\documentclass[10pt,a4paper]{article}
\usepackage[utf8]{inputenc}
\usepackage{amsmath}
\usepackage{amsfonts}
\usepackage{amssymb}
\usepackage{graphicx}
\usepackage{graphics}
\usepackage{epstopdf}
\usepackage{subfig}
\usepackage{float}
\usepackage{cite}
\graphicspath{{img/}}
\usepackage[toc,page]{appendix}
\usepackage{geometry}
\usepackage{authblk}
\usepackage{abstract}
\title{Analytical study of classic models of Hybrid Inflation}
\author[]{N. Malsawmtluangi \thanks{ Corresponding author, E-mail: \texttt{tei.naulak@uohyd.ac.in}}}
\affil[]{\small School of Physics, University of Hyderabad\\ Hyderabad-500046, India}

\date{}

\begin{document}
\maketitle

\begin{abstract}
We study the classic hybrid inflation model in its original and modified forms and show the shape of the inflationary potentials and analyze the amount of primordial gravitational waves each model predicts. We compare the resulting EE-mode and BB-mode power spectrum with the data from the joint BICEP2/Keck and Planck collaboration to check the viability of each model.
\end{abstract}


\section{Introduction}
The inflationary paradigm offers one of the most common scenarios of evolution of the early universe by suggesting the origins of density perturbations, which seeded the large scale structure of the universe, and tensor perturbations, which sourced primordial gravitational waves \cite{ag, al1, rhb, aas, ahg}. Both these perturbations are believed to have significant contributions to the cosmic microwave background anisotropies \cite{cmb1, cmb2, cmb3}.

Many models of inflation have been proposed and studied extensively \cite{enc}. Single scalar field models predict  an almost Gaussian distribution of temperature and density perturbations while models with multiple fields predict non-adiabatic and isocurvature perturbations and hence non-Gaussian distribution. However, observations of the CMB anisotropy indicate a negligible amount of non-Gaussianity and severe constraints on the amplitude of the isocurvature perturbations due to the anisotropy level of the CMB \cite{par1}. These results tend to favor single scalar field models. On the other hand, Planck and WMAP have observed a $\sim 10 \%$ hemispherical asymmetry in the power level of CMB at large angular scales \cite{hke, fkh, par2}.  This feature attracts several explanations and suggestions, one of which is that it could originate from inflation due to multiple fields with non-adiabatic and non-linear isocurvature perturbations from inflation \cite{lit, qy, iso, asm}.

A consistent feature in single and multiple field models is the existence of primordial gravitational waves. However, these waves are very weak and at present, their direct detection is deemed impossible. One way to realize their existence is through the imprint that they have left in the CMB anisotropies called the B-modes which can also be reflected on the BB-correlation angular spectrum of the CMB. The 2015 BICEP2/Keck and Planck collaboration data sets the bounds on tensor-to-scalar ratio as $r<0.07$ for the upper limit and the lower limit $r \simeq \mathcal{O}(10^{-3})$ \cite{par3, par4}.

Hybrid inflation model is a classic example of a multiple-field inflationary model. Since it generally predicts a blue tilt of scalar spectrum, it is often disfavored in its original form. However, it has, through the years, undergone several modifications which can produce agreeable results with observations. Moreover, hybrid inflationary models can be easily embedded in the frameworks of GUTs, supersymmetry, supergravity and string theories. Many models have been proposed and explored in these contexts \cite{gu1, gu2, gu3, ss1, ss2, ss3, ss4, sg1, sg2, sg3, stg}.

In this paper, we revisit the classic Hybrid Inflation model and some of its modifications \cite{al2, sfk, wfl, inv, mut, gl1, gl2, gl3}. We analyze the slow-roll inflationary potentials and calculate their tensor power spectrum and slow-roll parameters. We also plot the EE- and BB-mode power spectrum from the scalar and tensor modes which we have compared with the 2015 BICEP2/Keck and Planck collaboration data on the limit of the power spectrum.

\section{Inflationary Scenario}
Consider an inflation driven by $n$ scalar fields $\Phi_i$, where $i=1,2,...n$. Then, the equation of motion can be given by $n$ Klein-Gordon equations as \cite{mf1,mf2},
\begin{equation}
\ddot{\Phi}_i + 3H \dot{\Phi}_i + \frac{\partial V_i}{\partial \Phi_i} =0,
\end{equation}
where dot indicates derivative with respect to cosmic time $t$. $V_i=\sum_{i} V(\Phi_i)$ is the potential energy which is the sum of several terms and $H$ is the Hubble parameter and is determined by the energy density of the scalar field, $\rho_{\Phi_i} = \sum_{i} \frac{\dot{\Phi}_i^2}{2}+V_i,$ so that the Friedmann equation can be written as
\begin{equation}\label{fdmeq}
H^2 = \frac{1}{3 m^2_{pl}}\left( \frac{1}{2} \sum_{i=1}^{n} \dot{\Phi_i}^2 + V_i \right),
\end{equation}
where $m_{pl}$ is the reduced Planck mass.  The Hubble parameter in the above equation is due to the sum over all fields $\Phi_i$.

Let us introduce an inflaton field $\phi$ given by
\begin{equation}
\phi = \int \sum_{i=1}^{n}\hat{\phi}_i\dot{\Phi}_i dt
\end{equation}
that describes the evolution of all the fields along the direction given by the unit vector
\begin{equation}
\hat{\phi}_i = \frac{\dot{\phi}}{\sqrt{\sum_{i=1}^{n}\dot{\Phi}_i^2}}.
\end{equation}
Then, the equations of evolution of $n$ homogeneous scalar fields can be written as
\begin{equation}
\ddot{\phi} + 3H\dot{\phi} + V'=0,
\end{equation}
where $V'$ is the potential gradient in the field direction,
\begin{equation}
V' \equiv \frac{\partial V}{\partial \phi} = \sum_{i=1}^{n} \hat{\phi}_i \frac{\partial V}{\partial \Phi_i}.
\end{equation}
Then, the total energy density is given by that of the usual scalar field density $\rho_{\phi} = \frac{\dot{\phi}^2}{2}+V$. In the slow-roll limit, the energy density of the field is dominated by its potential energy such that $\frac{\dot{\phi}^2}{2} \ll V$. Hence,
\begin{equation}
H^2 \simeq \frac{V}{3m^2_{pl}},
\end{equation}
This slow-roll condition is characterized by the slow-roll parameters which can be defined in terms of the inflaton potential ($V$) and its derivatives as 
\begin{eqnarray}
\epsilon &\equiv & \frac{m_{pl}^2}{2}\left(\frac{V'}{V}\right)^2, \nonumber \\ \eta &\equiv & m^2_{pl}\left(\frac{V''}{V}\right),
\end{eqnarray}
and so on. Inflation lasts as long as the slow-roll conditions are satisfied, i.e., $\epsilon \ll 1$ and $|\eta|\ll 1$. In most scenarios, inflation ends by slow-roll violation followed by decay of the inflaton and then reheating which is followed by particle production.

The duration of inflation is characterized by the e-folding number $N$ which can be written in terms of the inflaton potential as
\begin{equation}
N \simeq \int_{\phi_{end}}^{\phi}\frac{V}{V'}d\phi
\end{equation}
where $\phi$ here corresponds to the value of the inflaton at horizon-crossing and $\phi_{end}$ corresponds to its value when the slow-roll limit becomes invalid. Throughout this paper, we take the e-folding value as 60.

In the slow-roll approximation, the power spectra of the scalar perturbation ($P_S$) and the tensor perturbation ($P_T$) generated outside the horizon can be given in terms of the potential by
\begin{eqnarray}
P_S &\simeq & \frac{1}{12 \pi^2 m^6_{pl}}\frac{V^3}{V'^2} \Big{|}_{k=aH},\\
P_T &\simeq & \frac{2}{3 \pi^2 m^4_{pl}}V \Big{|}_{k=aH},
\end{eqnarray}
where $k=aH$ indicates that both $H$ and $V$ are evaluated at the time when the mode with wave number $k$ crosses the horizon. Throughout this paper, we take the scalar power spectrum to be $P_S= 2.46 \times 10^{-9}$.

The tensor-to-scalar ratio can be written in terms of the equation of state parameter $\epsilon$ as
\begin{equation}\label{r}
r \equiv \frac{P_T (k)}{P_S (k)} \simeq 16\epsilon.
\end{equation}
This parameter measures the strength of the tensor perturbations relative to that of the scalar perturbations and the tensor spectral index is given by $n_T=-2\epsilon$. These parameters are therefore determined by the equation of state during inflation.

\section{Hybrid inflation models}
In this section, we discuss the classic Hybrid Inflation model and some of its modified forms and analyze their potentials, slow-roll conditions and their power spectrum. In the most general case, the inflation is driven by the inflaton field $\phi$ which slowly rolls down its potential while the non-inflationary field $\sigma$ initially remains at rest or almost constant at its false vacuum and governs the end of inflation due to symmetry breaking and is often called the waterfall field.

\subsection{Valley Hybrid Inflation model}
\begin{figure}[t]
\begin{center}
\includegraphics[scale=0.28]{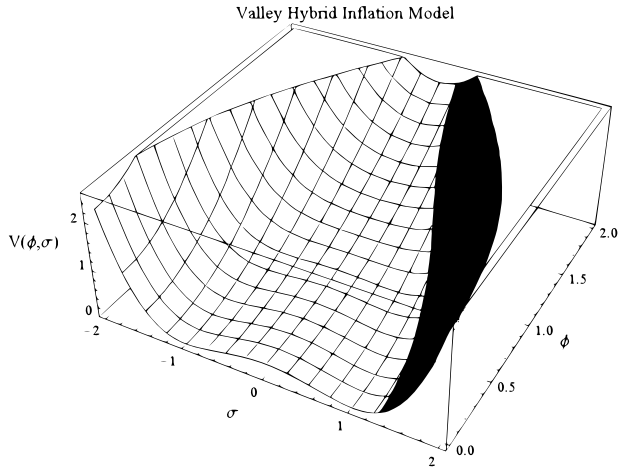}
\end{center}
\caption{Inflationary potential as a function of $\phi$ and $\sigma$ for valley hybrid inflation.}
\label{vhi}
\end{figure}
This is the classic hybrid inflation model where the inflation is driven by the inflaton field $\phi$ and its potential is given by \cite{ enc, al2}
\begin{equation}
V(\phi, \sigma) = \frac{\lambda_\sigma}{4}(\sigma^2-M^2)^2 + \frac{1}{2}m_\phi^2\phi^2 + \frac{1}{2}\lambda \phi^2\sigma^2,
\end{equation}
where $\lambda_\sigma$, $\lambda$ are the dimensionless coupling constants and $m_\phi=1.5\times 10^{-7} m_{pl}$ is the mass of the inflaton. $M$ is a one-dimensional constant and $\lambda_\sigma M^2=m^2_\sigma$ gives the mass of the non-inflationary scalar $\sigma$.

The form of the potential for this model is shown in figure \ref{vhi}. In this model, the $\sigma$ field is initially at the origin due to its interaction with the inflaton field $\phi$. The $\phi$ field starts at a value greater than the critical value $\phi_c = m_\sigma / \sqrt{\lambda}$ and slowly rolls down along the valley given by $\sigma=0$, and when it reaches the origin, it displaces the $\sigma$ field from its minimum potential. This eventually results in symmetry breaking and fast roll thus bringing about an abrupt end to inflation.

Define the parameters \[\Lambda = \frac{\lambda_\sigma^{1/4}M}{\sqrt{2}},~~~~~~~~~~{\rm and}~~~~~~~~~~\mu=\sqrt{\frac{\lambda_\sigma}{2}}\frac{M^2}{m_\phi}.\]
Then, when $\sigma=0$, the potential reduces to
\begin{equation}
V=\Lambda^4\left(1+\frac{1}{2}\frac{\phi^2}{\mu^2}\right).
\end{equation}
For this model, the slow-roll parameters are
\begin{eqnarray}
\epsilon &=& 2.6 \times 10^{-4},  \\
\eta &=& 1.5 \times 10^{-4}.
\end{eqnarray}
This leads to the tensor spectral index $n_T = -5.2 \times 10^{-4}$. Tensor power spectrum is $P_T = 1.03 \times 10^{-11}$. This gives the tensor-to-scalar ratio, $r=4.2 \times 10^{-3}$.

\subsection{Inverted Hybrid Inflation model}
\begin{figure}
\begin{center}
\includegraphics[scale=0.28]{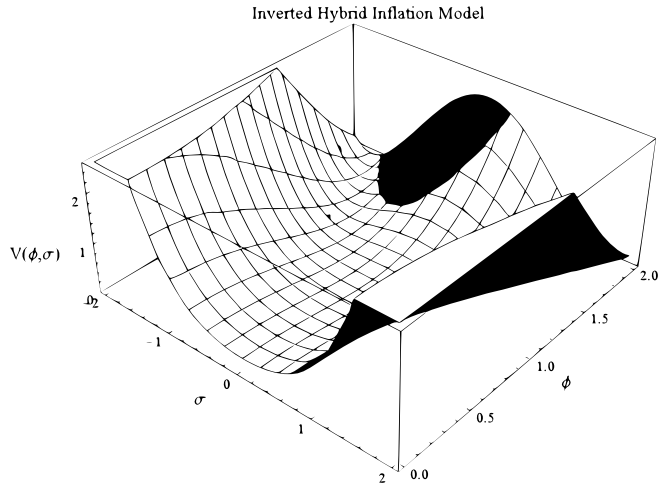}
\end{center}
\caption{Inflationary potential as a function of $\phi$ and $\sigma$ for inverted hybrid inflation}
\label{ihi}
\end{figure}
In this model, the $\phi$ field rolls away from this origin and its potential is obtained by just reversing certain signs in the potential of valley hybrid model (VHI).
 However, the inflaton field is supposed to obtain a vacuum expectation value eventually, and this is achieved by adding a quartic term to the potential \cite{sfk, inv}
\begin{equation}
V(\phi, \sigma) = \frac{\lambda_\sigma}{4}(\sigma^2+M^2)^2 - \frac{1}{2}m_\phi^2\phi^2 - \frac{1}{2}\lambda \phi^2\sigma^2 + \frac{1}{4}\lambda_\phi \phi^4,
\end{equation}
where we have added a quartic term $\phi^4$ and its coupling constant $\lambda_\phi$ in the potential. $m_\phi$ is the mass of a soft supersymmetry-breaking mass term $\phi$ and is of the order of $\sim 1$ TeV. The constant $M$ is of the order of $10^{11}$ GeV.

The potential for this inflation model is shown in figure \ref{ihi} where the path of the inflaton is away from the origin. When the $\phi$ field is rolling down its potential, its interaction with the $\sigma$ field holds the latter in its own false vacuum $\sigma=0$.
\begin{equation}
V(\phi) = \frac{\lambda_\sigma M^4}{4} - \frac{1}{2}m_\phi^2\phi^2 + \frac{1}{4}\lambda_\phi \phi^4.
\end{equation}
 When $\phi$  reaches some critical value $\phi_c = m_\sigma / \sqrt{\lambda}$, it releases its hold on the $\sigma$ field which undergoes second order phase transition and obtains its vacuum expectation value.

In this model, inflation occurs when $\phi < \phi_c$. When $\sigma=0$, the inflaton potential is minimized at $\phi_{min}=m/\sqrt{\lambda_\phi}$. In the hybrid mechanism, $\phi_c$ must be lower than $\phi_{min}$, since the $\sigma$ field remains trapped in its false vacuum until $\phi$ reaches $\phi_c$. If $\phi_c > \phi_{min}$, the $\sigma$ field would never be able to achieve its true vacuum and the inflationary scenario would be similar to the usual single-field driven one. 

If we impose the condition $\phi_c \ll \phi_{min}$, we can neglect the quartic term and we obtain the potential
\begin{equation}
V=\Lambda^4\left(1-\frac{1}{2}\frac{\phi^2}{\mu^2}\right),
\end{equation}
where the parameters $\Lambda$ and $\mu$ hold the same definition as in the case of VHI. However, the dimensionless parameters are very small as compared to those in VHI in order to satisfy the condition $\phi_c < \phi_{min}$, 
\begin{equation}
\lambda_\phi \ll \left(\frac{m_\phi}{\phi_c}\right)^2,
\end{equation}
and $\lambda$ and $\lambda_\sigma$ are even smaller than $\lambda_\phi$.

For this model, the slow-roll parameters are
\begin{eqnarray}
\epsilon &=& 3.3 \times 10^{-5},  \\
\eta &=& -1.49 \times 10^{-2}.
\end{eqnarray}
This leads to the tensor spectral index $n_T = -6.6 \times 10^{-5}$. Tensor power spectrum is $P_T = 1.3 \times 10^{-12}$. This gives the tensor-to-scalar ratio, $r=5.3 \times 10^{-4}$, which is very small.

\subsection{Mutated Hybrid Inflation model}
\begin{figure}
\begin{center}
\includegraphics[scale=0.28]{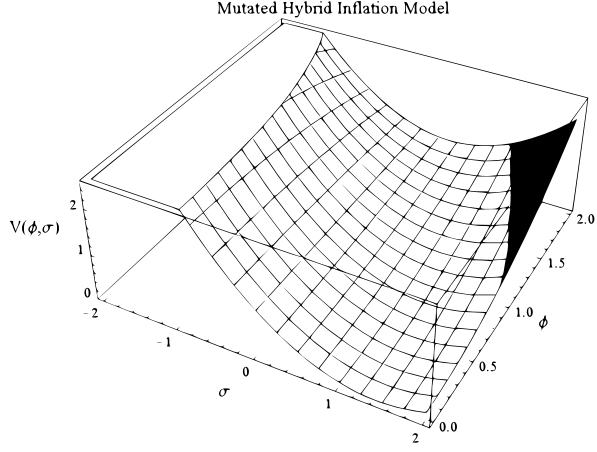}
\end{center}
\caption{Inflationary potential as a function of $\phi$ and $\sigma$ for mutated hybrid inflation}
\label{mhi}
\end{figure} 
In mutated hybrid inflation model, the scalar fields $\phi$ and $\sigma$ possess canonical kinetic terms and the effective potential takes the form \cite{inv, mut}
\begin{equation}
V(\phi, \sigma)= \frac{1}{2}m^2(\sigma-\sqrt{2}M)^2+\frac{1}{4}\lambda \phi^2 \sigma,
\end{equation}
where it is assumed that $m \ll \lambda \lesssim 1$ and $M \lesssim 1$.

The potential for mutated hybrid inflation model is shown in figure \ref{mhi}. In this model, the inflation is driven by the inflaton $\phi$ while the $\sigma$ field is held close to zero, but not actually at zero, $\sigma \neq 0$ such that $\phi > 0$, $\sigma > 0$.

 Define a parameter
\begin{equation}
\alpha(\phi)=\frac{2m^2}{\lambda^2 \phi^2},
\end{equation}
such that inflation occurs for $\phi^2 \gg \alpha$, then, we can write
\begin{equation}
V(\phi, \sigma)=\frac{m^2M^2}{1+\alpha}+\frac{1}{4}(1+\alpha)\lambda^2\phi^2 \left(\sigma-\frac{\sqrt{2}\alpha M}{1+\alpha}\right)^2.
\end{equation}
For fixed $\phi$, the model has a minimum at $\sigma=\sigma_\ast >0$. Thus, $V_\sigma =0$ when $\sigma=0$. Taking the first derivative w.r.t $\sigma$,
\begin{equation}
V_{\sigma}=\frac{1}{2}(1+\alpha)\lambda^2\phi^2 \left(\sigma-\frac{\sqrt{2}\alpha M}{1+\alpha}\right).
\end{equation}
Therefore, we have
\begin{equation}\label{ab}
\sigma = \sigma_\ast \equiv \frac{\sqrt{2}\alpha M}{1+\alpha} \simeq \frac{2\sqrt{2}m^2M}{\lambda^2\phi^2}.
\end{equation}
Thus, if $\phi > \sigma$ initially, the $\sigma$ field will decrease much more rapidly than the inflaton. As such, the fields will rapidly approach the inflationary trajectory $\sigma \phi^2$ given by (\ref{ab}) with $\phi^2 \gg m/\lambda \gg \sigma$. Also, it is assumed that $\phi \ll 1$, since higher order terms in $\phi$ become prominent otherwise.

Thus on the assumption that $\sigma$ is constrained at $\sigma=\sigma_\ast$ during inflation, we get the inflationary potential,
\begin{equation}
V\Big{|}_{\sigma=\sigma_\ast}=\frac{m^2M^2}{1+\alpha} \simeq m^2M^2 \left(1-\frac{2m^2}{\lambda^2\phi^2}\right),
\end{equation}
and the non-minimal kinetic terms evaluated along $\sigma=\sigma_\ast$,
\begin{equation}
\frac{1}{2}\left(1+\frac{8\alpha^2 M^2}{\phi^2}\right)(\partial \phi )^2
\end{equation}
in which, since $\phi^2 \gg \sigma$ during inflation, $\sqrt{2}\alpha M \ll \phi$, hence the second term in the brackets is very small and the kinetic terms are approximately canonical and can be neglected.

Let $m^2M^2=\Lambda^4$ and $2m^2/\lambda^2=\mu$. Then, the potential becomes
\begin{equation}
V=\Lambda^4\left(1-\frac{\mu}{\phi^2}\right).
\end{equation}
The e-folding number is
\begin{eqnarray}
N \simeq \frac{\phi^4}{8\mu}.
\end{eqnarray}
This leads to 
\begin{equation}
m = \frac{\lambda \phi^2}{4\sqrt{N}}.
\end{equation}
Since $\phi \ll 1$, we have 
\begin{equation}
m \ll \frac{\lambda}{4\sqrt{N}}.
\end{equation}
For this model, the slow-roll parameters are
\begin{eqnarray}
\epsilon &=& 3.91 \times 10^{-4},  \\
\eta &=& 6.71 \times 10^{-4}.
\end{eqnarray}
This leads to the tensor spectral index $n_T = -7.8 \times 10^{-4}$. Tensor power spectrum is $P_T = 1.52 \times 10^{-11}$. This gives the tensor-to-scalar ratio, $r=6.25 \times 10^{-3}$.

\subsection{Smooth Hybrid Inflation model}
\begin{figure}
\begin{center}
\includegraphics[scale=0.28]{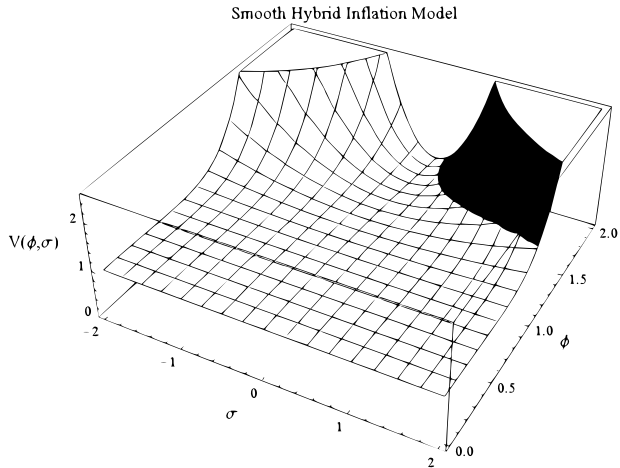}
\end{center}
\caption{Inflationary potential as a function of $\phi$ and $\sigma$ for smooth hybrid inflation}
\label{shi}
\end{figure}
This model is applicable in a wide class of SUSY GUT models. However, we shall skip the initial superpotential and the transformation of its complex singlet superfields to the two real scalar fields and directly write the potential which is of the form \cite{gl1, gl2, gl3}
\begin{equation}\label{b}
V(\phi, \sigma)=\left(\mu^2-\frac{\phi^4}{16M^2}\right)^2+\frac{\phi^6\sigma^2}{16M^4},
\end{equation}
with the supersymmetric minima corresponding to
\begin{eqnarray}\label{vac}
|< \phi > | &=& 2(\mu M)^{1/2}, \nonumber \\
< \sigma > &=& 0,
\end{eqnarray}
where $\mu$ is a superheavy mass scale %related to grand unification scale 
and $M$ is mass scale of the order of compactification scale $\sim 10^{18}$ GeV.

The form of the inflationary potential for this model is shown in figure \ref{shi} which appears smooth compared to other models. In this model, throughout the inflationary process, the system follows a particular path along the valley of minima which leads to a particular point of the vacuum manifold obtained from (\ref{vac}).  As such, the end of inflation is not abrupt unlike the previous cases and is rather smooth, and there is no topological defect.

For any fixed $\sigma$, the potential $V(\phi)$ has local maximum at $\phi^2=0$ and absolute minimum at 
\begin{equation}\label{min}
\phi^2 \simeq \frac{4}{3} \frac{\mu^2 M^2}{\sigma^2},~~~~~~~~~~\forall ~ \sigma^2 \gg \mu M.
\end{equation}
The $\phi$ field performs damped oscillations initially over its local maximum while the $\sigma$ field remains approximately constant. Within an interval of time given by $\Delta t \sim 6\pi(\sigma /m_{pl})^2H^{-1}$, the $\phi$ field falls into the valley of the minima given in (\ref{min}) and settles at the bottom of this valley while the $\sigma$ field remains approximately constant and $\sigma \gg m_{pl}$.

The potential along the maxima at $\phi^2=0$ is constant and is equal to $\mu^4$ while along the valley of minima, the potential takes the form
\begin{equation}
V_{min}(\sigma) \simeq \Lambda^4 \left(1-\frac{2}{27}\frac{\mu^2 M^2}{\sigma^4}\right),
\end{equation}
where $\Lambda = \mu$ is the scale of inflation  
where $\mu \simeq 8.7 \times 10^{14}$ GeV and $M \simeq 9.4 \times 10^{17}$ GeV using COBE normalization \cite{gl1, arl}.

After the end of inflation, the $\phi$ and $\sigma$ fields enter an oscillatory phase smoothly about the global supersymmetric minima in (\ref{vac}) and eventually decay into lighter particles, and thus subsequently bringing about the reheating of the universe.

For this model, the slow-roll parameters are evaluated on the valley of minima of (\ref{min}). The results obtained are
\begin{eqnarray}
\epsilon &=& 5.62 \times 10^{-8},  \\
\eta &=& -1.51 \times 10^{-2}.
\end{eqnarray}
This leads to the tensor spectral index $n_T = -1.12 \times 10^{-7}$. Tensor power spectrum is $P_T = 2.2 \times 10^{-15}$. This gives the tensor-to-scalar ratio, $r=8.9 \times 10^{-7}$ which is very small and negligible compared to the scalar perturbations.

\section{Power Spectrum}
The  $BB$-mode correlation  angular power spectrum of  CMB is given by \cite{cmb1, cmb2}
\begin{eqnarray}
C_l^{BB} &=& (4\pi)^2 \int dk k^2 P_T(k) \nonumber \\ && \times \left| \int_0^{\tau_0} d\tau g(\tau) h_k(\tau) \Big\{(8x + 2x^2 \partial_x)\frac{j_l(x)}{x^2}\Big\}_{x=k(\tau_0-\tau)}\right|^2,
\end{eqnarray}
where $g(\tau) = \kappa e^{-\kappa}$ is the probability distribution of the last scattering, $\kappa$ is the differential optical depth, $x= k(\tau_0 -\tau)$ and $j_l(x)$ is the spherical Bessel function.

The tensor power spectrum $P_T(k)$ for different inflation models  depends  on the tensor spectral index $n_T$, which varies according to the effective potential of each model.

After calculations of the necessary parameters, data for angular spectra for the aforementioned inflation models are generated using the CAMB code with the parameters corresponding to each model.  For all models, the optical depth is taken to be $\kappa = 0.09$, the pivot wave number for tensor mode is taken as $k_0 =0.002 ~{\rm Mpc^{-1}}$ and that for scalar mode is  $k_0 =0.05~{\rm Mpc^{-1}} $. Note that for each plot, the quantity plotted against the multipole ($l$) is $l(l+1)C_l/2\pi$ in units of $\mu K^2$, where $C_l$ is the $l$-th amplitude in the power spectrum.
\begin{figure}[H]
\begin{center}
\includegraphics[scale=0.32]{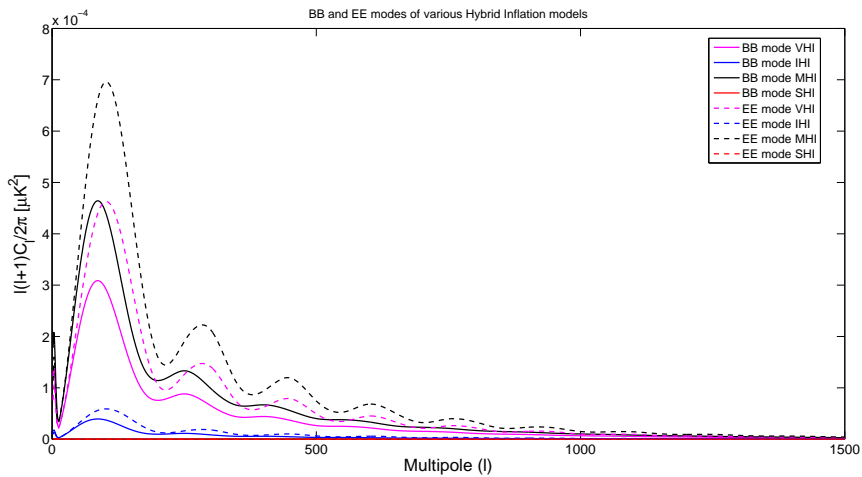}
\end{center}
\caption{EE- and BB-mode spectra from tensor mode.}
\label{be_all}
\end{figure}
In figure \ref{be_all}, we show the BB- and EE-modes of various hybrid inflation models induced by tensor mode. The solid lines indicate the BB-modes while the dashed lines indicate the EE-modes. The magenta lines represent the spectra for valley hybrid inflation model (VHI), the blue lines indicate inverted hybrid inflation model (IHI), the black lines indicate mutated hybrid inflation model (MHI) and the red lines represent smooth hybrid inflation model (SHI). It can be seen that auto-correlation spectrum of the scalar perturbations have much higher power spectrum than those of tensor perturbations for each model, as already known well. Also, the EE- and BB-spectra for SHI have very low power compared to other models.
\begin{figure}[H]
\centering
\subfloat[]
{\includegraphics[scale=0.18]{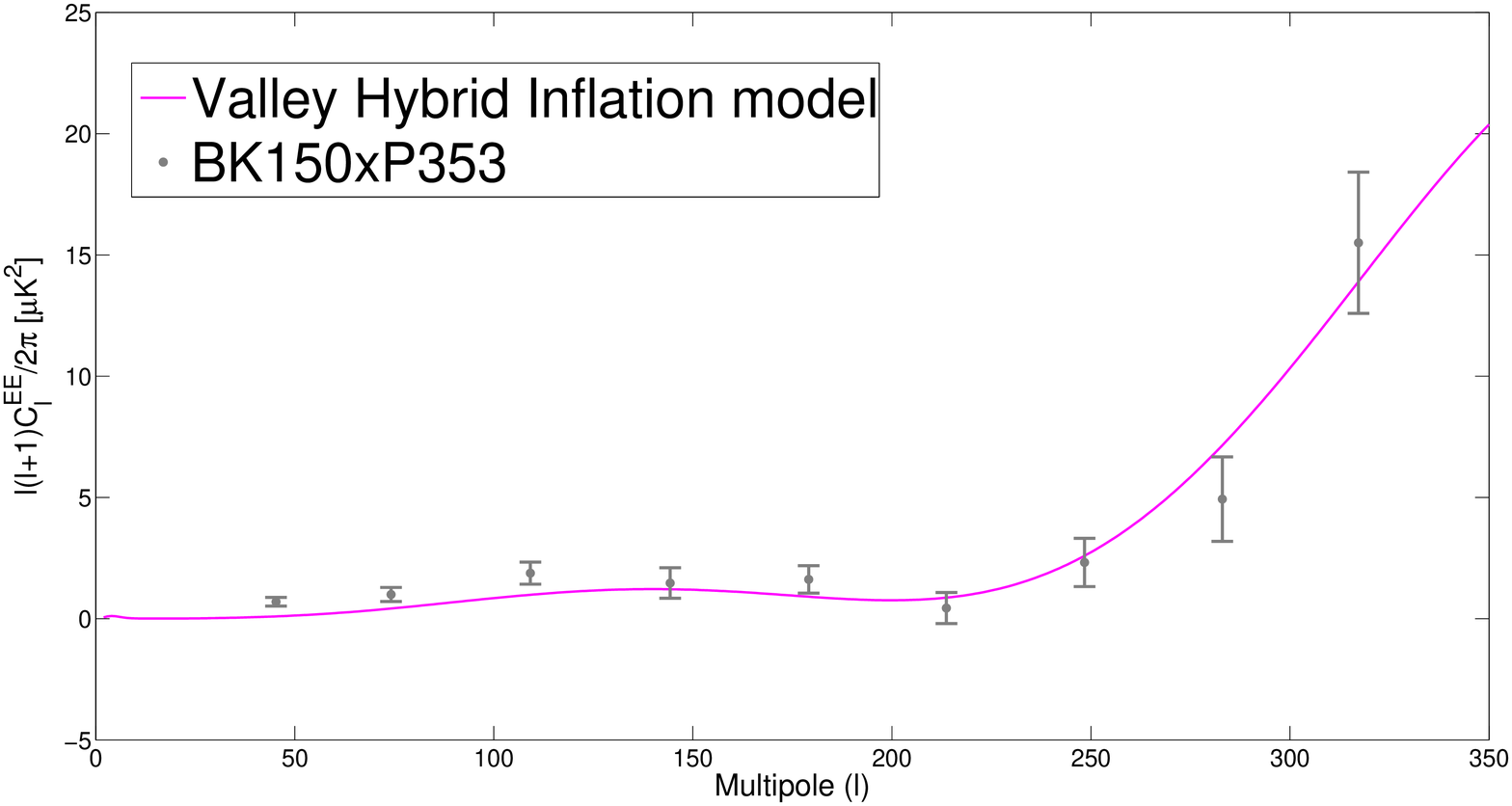}
\label{ee_v}}
\hfill
\subfloat[]
{\includegraphics[scale=0.18]{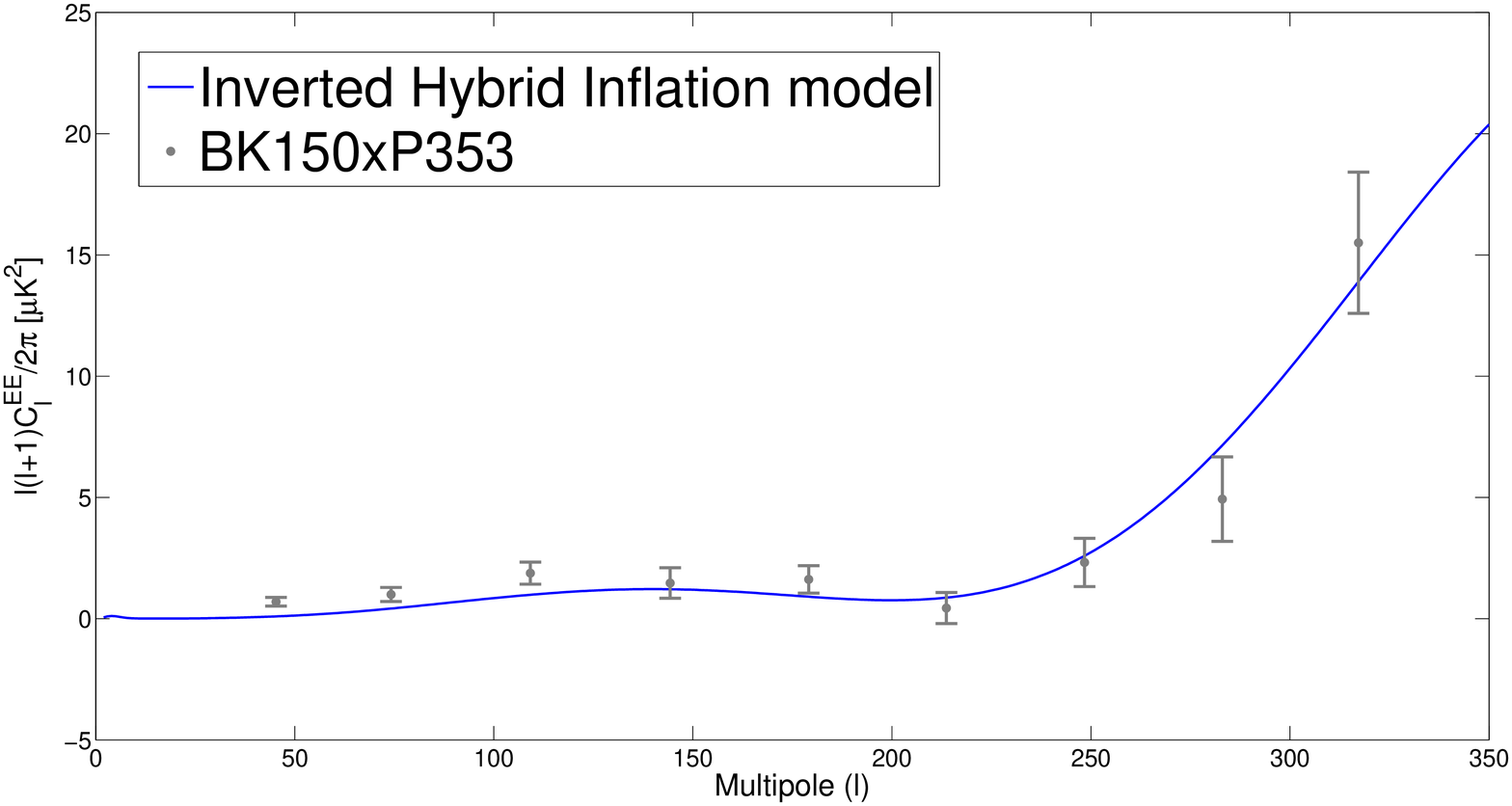}
\label{ee_i}}
\hfill
\subfloat[]
{\includegraphics[scale=0.18]{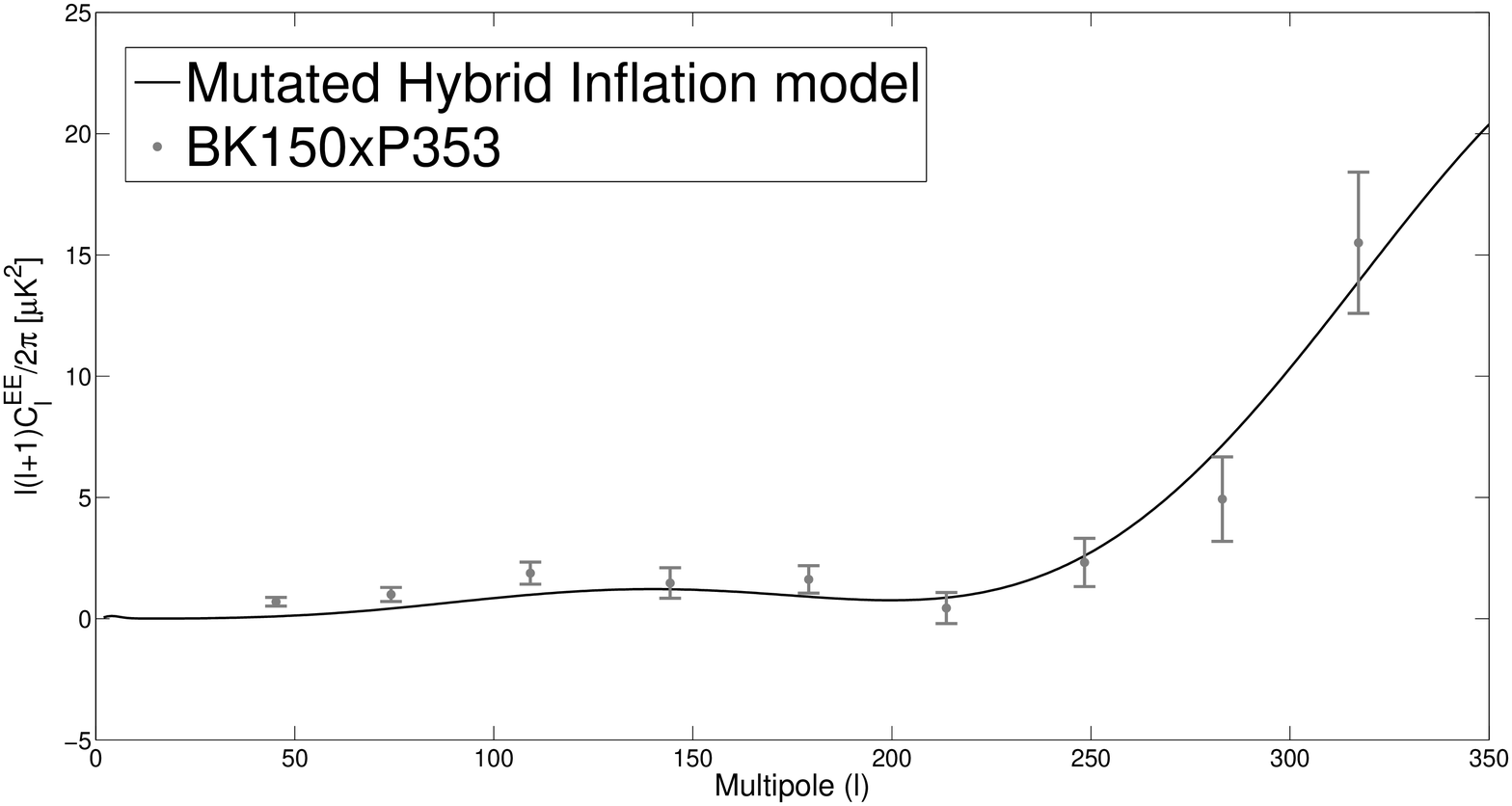}
\label{ee_m}}
\hfill
\subfloat[]
{\includegraphics[scale=0.18]{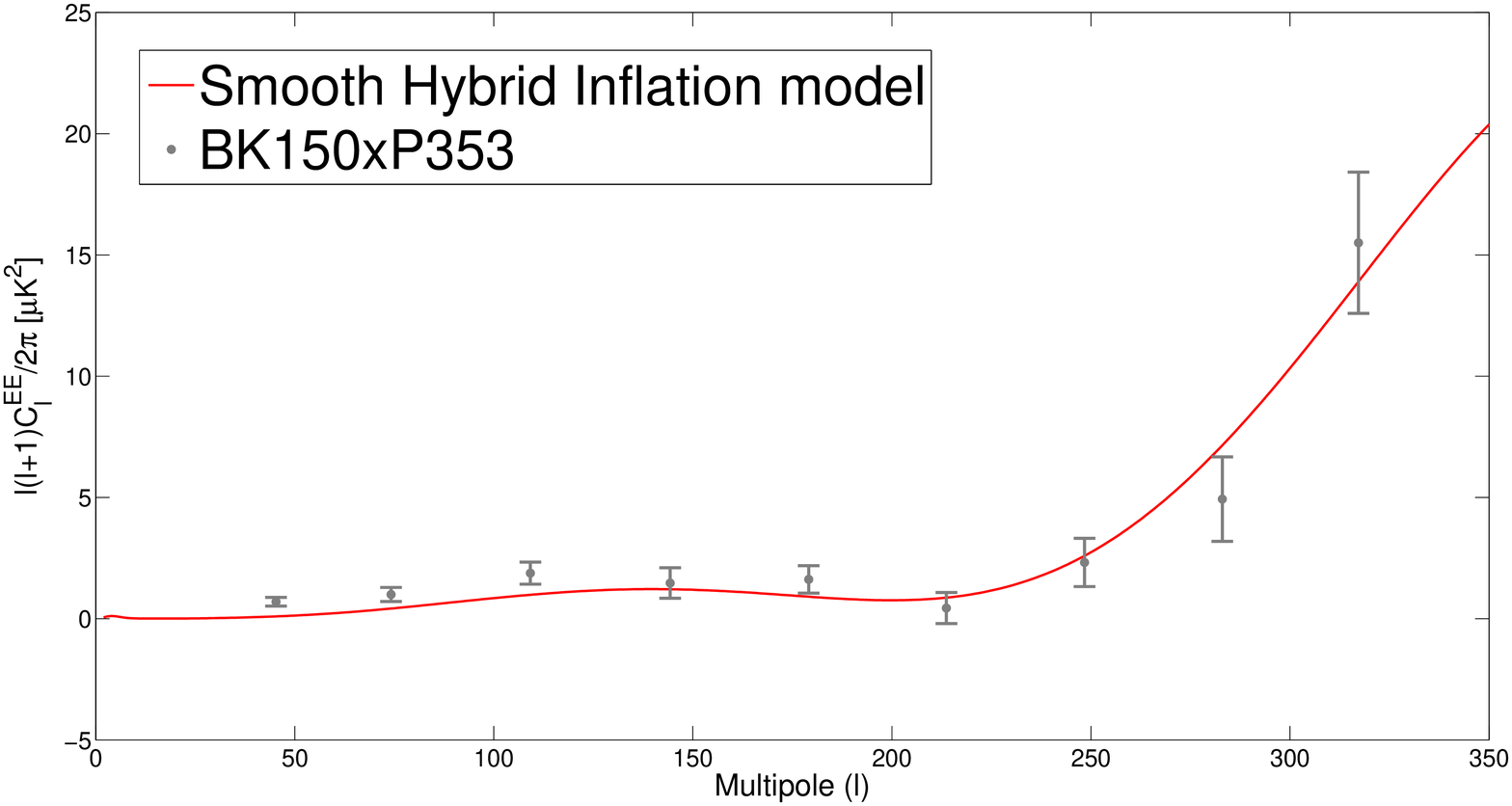}
\label{ee_s}}
\caption{EE-modes from scalar mode compared with BICEP2/Keck+Planck errorbar.}
\label{ee_models}
\end{figure}
Figures \ref{ee_models} show the EE-mode spectrum generated by the scalar mode for each model compared with BK150xP353 data. The errorbars are the standard deviations of lensed-$\Lambda$CDM$+$noise simulations. For each model, the EE-modes are more or less in line with the bound, the enhancement of the BKP errorbar at around $l \sim 100$ is due to the weak dust contribution. %Note that the lensed effects are incorporated in these plots in accordance with the available BKP data.
\begin{figure}[H]
\begin{center}
\includegraphics[scale=0.37]{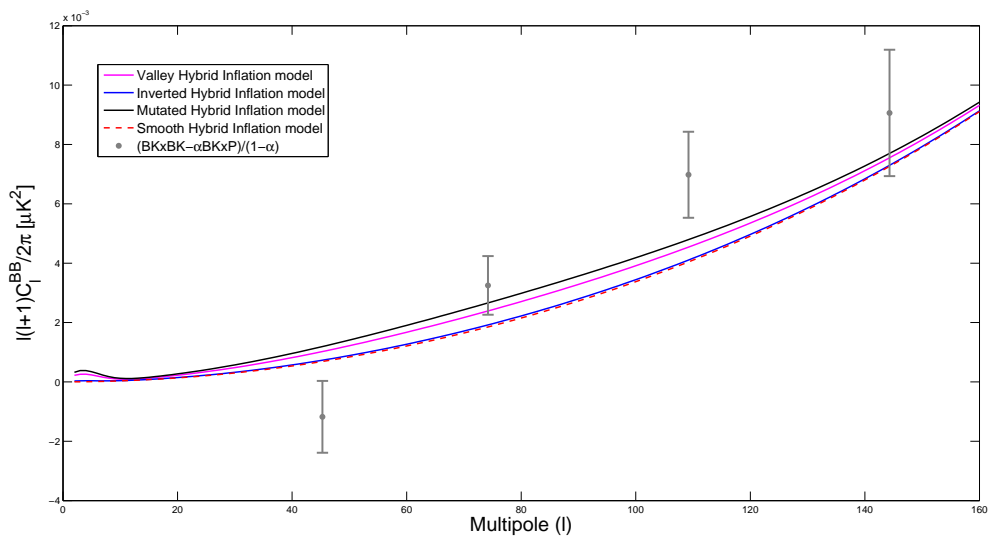}
\end{center}
\caption{Lensed BB-mode spectrum compared with BICEP2/Keck+Planck errorbar.}
\label{ft_c_all}
\end{figure}
What we can see from figures \ref{be_all} and \ref{ee_models} regarding the EE-modes is that the EE-modes from tensor perturbations have drastic difference in power level for various inflation models. However, the EE-modes from scalar perturbations have very little difference in power level, which can be seen from figure \ref{ee_models}. Moreover, the EE-modes from scalar mode have much higher power level than those from the tensor modes.

The results for the lensed BB-modes are also compared with the data from the joint analysis of BK150xPlanck353 as shown in figure \ref{ft_c_all}. The limit (BKxBK - $\alpha$BKxP)/(1 - $\alpha$) at $\alpha = \alpha_{fid} = 0.04$ is evaluated from the auto-spectra of the combined BICEP2/Keck 150 GHz maps and its cross-spectra with Planck 353 GHz maps respectively. This combination for the limit is taken after the subtraction of the dust contribution which is 0.04 times as much in the BICEP2 band as it is in the Planck 353 GHz band.

 It can be seen from figure \ref{ft_c_all} that the valley hybrid and mutated hybrid inflation models have BB-mode spectrum which are just within the BKP bound. The inverted hybrid inflation model also predicts very small amount of primordial gravitational waves. Similarly, while the smooth hybrid inflation model has the nicest scenario, in the sense that it has no waterfall regime and hence, no topological defects post inflation, its tensor perturbations are utterly small compared to its scalar perturbations as can be seen from the calculations, and has the lowest power spectrum of the four hybrid models, which is very much out of bound. Its pure (unlensed) BB-mode spectrum is very low compared to the other models and  even looks flat even though it has basically the same shape as the others albeit at lower spectrum. Even though all the models are within the bound around multipole $l \sim 145$, the BB-mode spectrum at the higher multipoles are contaminated with lensed B-modes, which are actually E-modes (scalar perturbations) converted to B-modes at later times due to gravitational lensing, hence, the higher order multipoles are mostly neglected in the analysis.

\section{Conclusion}
The hybrid model and its modified forms are revisited and analyzed. In can be seen from figure \ref{ft_c_all} that the predicted power spectrum of BB-modes of valley hybrid inflation model and mutated inflation model are in good agreement with the BKP data. We have also shown in figures \ref{vhi}-\ref{shi} the shape of the inflationary potential for each model where the inflaton starts rolling slowly from the top. It can be seen where the inflaton continues to roll. In all the cases of valley and inverted hybrid models, the inflation ends with the waterfall regime. Topological defects are usually formed during waterfall phase transitions. 
If that is the case, the smooth hybrid inflation, even though it predicts negligible amount of primordial gravitational waves, has the most advantageous scenario.
However, it has been found that an e-folding of 60 during the waterfall regime could inflate away the topological defects \cite{wfl}.  In such cases, all the aforementioned models are still in good favor.

\section*{Acknowledgment}
The author thanks CAMB, Planck and BICEP2/Keck websites for necessary tool and data.

%\section*{References}

\end{document}